%% file: APV_main.tex
\documentclass[12pt]{article}
\input{Rosetta+}

\newcommand{\nll}{\nonumber \\}

\newcommand{\um}{m_u}
\newcommand{\ums}{\um^2}
\newcommand{\dm}{m_d}

\newcommand{\ruM}{r_{u\sss{M}}}

\newcommand{\ruw}{r_{u\sss{W}}}

\newcommand{\rdw}{r_{d\sss{W}}}
\newcommand{\rhz}{r_{ \sss{HZ}}}
\newcommand{\rhw}{r_{ \sss{HW}}}
\def\citenum#1{{\def\@cite##1##2{##1}\cite{#1}}}
\def\citea#1{\@cite{#1}{}}
\begin{document}
\input{APV_title}

\input{APV_text}

\input{expansions}
%-----------------

\input{bibl_APV}
\end{document}

%% file: Rosetta+.tex
\newcommand{\nl}{\nonumber\\}

\newcommand{\ds}{\displaystyle}

\newcommand{\lpar}{\left(}                            % bracketing
\newcommand{\rpar}{\right)} 
\newcommand{\lrbr}{\left[}
\newcommand{\rrbr}{\right]}
\newcommand{\lcbr}{\left\{}

\newcommand{\bq}{\begin{equation}}                    % equationing
\newcommand{\eq}{\end{equation}}
\newcommand{\bqa}{\begin{eqnarray}}
\newcommand{\eqa}{\end{eqnarray}}
\newcommand{\ba}[1]{\begin{array}{#1}}
\newcommand{\ea}{\end{array}}
\newcommand{\ben}{\begin{enumerate}}
\newcommand{\een}{\end{enumerate}}
\newcommand{\bei}{\begin{itemize}}
\newcommand{\eei}{\end{itemize}}
\newcommand{\eqn}[1]{Eq.(\ref{#1})}

\newcommand{\eqnsc}[2]{Eqs.(\ref{#1}) and (\ref{#2})}

\newcommand{\tabn}[1]{Tab.~\ref{#1}}

\newcommand{\MeV}{\mathrm{MeV}}

\def\Re{\mathop{\operator@font Re}\nolimits}
\def\Im{\mathop{\operator@font Im}\nolimits}
\newcommand{\ord}[1]{{\cal O}\lpar#1\rpar}

\newcommand{\asums}[1]{\sum_{#1}}

\newcommand{\fer}{{\rm{fer}}}
\newcommand{\bos}{{\rm{bos}}}

\newcommand{\had}{{h}}

\newcommand{\ph}{\gamma}

\newcommand{\wb}{W}

\newcommand{\zb}{Z}

\newcommand{\fe}{e}
\newcommand{\ff}{f}

\newcommand{\barf}{\overline f}

\newcommand{\mw}{M_{_W}}

\newcommand{\mz}{M_{_Z}}

\newcommand{\mh}{M_{_H}}

\newcommand{\mt}{m_t}

\newcommand{\muq}{m_u}
\newcommand{\md}{m_d}

\newcommand{\ms}{m_s}

\newcommand{\mws}{M^2_{_W}}

\newcommand{\mzs}{M^2_{_Z}}

\newcommand{\mhs}{M^2_{_H}}
\newcommand{\mfs}{m^2_f}

\newcommand{\mts}{m^2_t}
\newcommand{\mes}{m^2_e}

\newcommand{\minds}[1]{m^2_{#1}}
\newcommand{\sman}{s}
\newcommand{\tman}{t}

\newcommand{\als}{\alpha_{_S}}

\newcommand{\gf}{G_{\ssF}}

\newcommand{\hatv}[1]{{\hat v}_{#1}}

\newcommand{\siws}{s^2_{_W}}
\newcommand{\cows}{c^2_{_W}}

\newcommand{\siwf}{s^4_{_W}}
\newcommand{\cowf}{c^4_{_W}}

\newcommand{\shs}{{\hat s}^2}
\newcommand{\chs}{{\hat c}^2}

\newcommand{\gfd}{\gamma_5}

\newcommand{\gdp}{\gamma_+}

\newcommand{\gadu}[1]{\gamma_{#1}}

\newcommand{\gz}{\Gamma_{_{\zb}}}

\newcommand{\tcie}{I^{(3)}_{\fe}}

\newcommand{\tcif}{I^{(3)}_{\ff}}

\newcommand{\qu}{Q_u}

\newcommand{\qf}{Q_f}

\newcommand{\Sww}{\Sigma_{_{\wb\wb}}}

\newcommand{\Szz}{\Sigma_{_{\zb\zb}}}

\newcommand{\MSB}{\overline{MS}}

\newcommand{\gpar}{\xi}

\newcommand{\sbff}[1]{B_{#1}}                    % B form-factors
\newcommand{\sfbff}[1]{B^{\ssF}_{#1}}
\newcommand{\bff}[4]{B_{#1}\lpar #2;#3,#4\rpar}             
             
\newcommand{\fbff}[4]{B^{\ssF}_{#1}\lpar #2;#3,#4\rpar}

\newcommand{\cff}[7]{C_{#1}\lpar #2,#3,#4;#5,#6,#7\rpar}    
 
\newcommand{\tHs}{\mu}

\newcommand{\tHss}{\mu^2}

\newcommand{\vverti}[3]{F^{#1}_{#2}\lpar{#3}\rpar}

\newcommand{\upar}[1]{u}

\newcommand{\ssF}{{\scriptscriptstyle{F}}}

\newcommand{\ssW}{{\scriptscriptstyle{W}}}

\newcommand{\bqas}{\begin{eqnarray*}}
\newcommand{\eqas}{\end{eqnarray*}}
\newcommand{\afba}[1]{A^{#1}_{_{\rm FB}}}

\def\app#1#2 {{\it Acta. Phys. Pol.} {\bf#1},#2}
\def\cpc#1#2 {{\it Computer Phys. Comm.} {\bf#1},#2}
\def\np#1#2 {{\it Nucl. Phys.} {\bf#1},#2}
\def\pl#1#2 {{\it Phys. Lett.} {\bf#1},#2}
\def\prep#1#2 {{\it Phys. Rep.} {\bf#1},#2}
\def\prev#1#2 {{\it Phys. Rev.} {\bf#1},#2}
\def\prl#1#2 {{\it Phys. Rev. Lett.} {\bf#1},#2}
\def\zp#1#2 {{\it Zeit. Phys.} {\bf#1},#2}
\def\sptp#1#2 {{\it Suppl. Prog. Theor. Phys.} {\bf#1},#2}
\def\mpl#1#2 {{\it Modern Phys. Lett.} {\bf#1},#2}
\def\jetp#1#2 {{\it Sov. Phys. JETP} {\bf#1},#2}
\def\fpj#1#2 {{\it Fortschr. Phys.} {\bf#1},#2}
\def\afp#1#2 {{\it Acta.Phys. Polon.} {\bf#1},#2}
\def\err#1#2 {{\it Erratum} {\bf#1},#2}
\def\ijmp#1#2 {{\it Int. J. Mod. Phys} {\bf#1},#2}
\def\nc#1#2 {{\it Nuovo Cimento} {\bf#1},#2}
\def\ap#1#2 {{\it Ann. Phys.} {\bf#1},#2}
\def\cmp#1#2 {{\it Comm. Math. Phys.} {\bf#1},#2}
\def\el#1#2 {{\it Europhys. Lett.} {\bf#1},#2}
\def\hpa#1#2 {{\it Helv. Phys. Acta} {\bf#1},#2}
\def\yf#1#2 {{\it Yad. Fiz.} {\bf#1},#2}
\def\nim#1#2 {{\it Nucl. Instrum. Meth.} {\bf#1},#2}
\def\spz#1#2 {{\it Sov. Pisma Zhetf} {\bf#1},#2}
\def\jetpl#1#2 {{\it JETP Lett.} {\bf#1},#2}
\def\sjnp#1#2 {{\it Sov. J. Nucl. Phys.} {\bf#1},#2}
\def\ptp#1#2 {{\it Progr. Theor. Phys. (Kyoto)} {\bf#1},#2}
\def\rmp#1#2  {{\it Rev. Mod. Phys.} {\bf#1},#2}
\def\zhetf#1#2 {{\it ZhETF} {\bf#1},#2}
\def\prs#1#2 {{\it Proc. Roy. Soc.} {\bf#1},#2}
\def\phys#1#2 {{\it Physica} {\bf#1},#2}

\newcommand{\sss}[1]{\scriptscriptstyle{#1}}

%% file: APV_title.tex
\thispagestyle{empty}

{\flushright 
\begin{tabular}{cr}
\hspace{12.2cm} &{\large February 19, 2001}\\
                &{\large hep-ph/0102233}
\end{tabular}
}

\vspace{2cm}

\begin{center}
{\LARGE\bf Atomic Parity-Violation and Precision Physics}

\vspace{2.5cm}

{\large
D. Bardin${}^1$, P. Christova${}^{1,2,\dagger}$,
L. Kalinovskaya${}^1$ and Giampiero Passarino${}^{3,\dagger}$} \\[2cm]
$^{1}$ {\it Laboratory of Nuclear Problems, JINR, Dubna, Russia} \\[3mm]
$^{2}$ {\it Faculty of Physics, Bishop Preslavsky University, Shoumen, Bulgaria}
   \\[3mm]
$^{3}$ {\it Dipartimento di Fisica Teorica, Universit\`a di Torino, Italy}
   \\[2mm]
   {\it INFN, Sezione di Torino, Italy} \\[10mm]
\vspace{2cm}
%\maketitle

{\large\bf Abstract}
 
\end{center}

%\begin{abstract}
  \normalsize \noindent 
The atomic parity-violation (APV) parameter $Q_{\sss{W}}$ for
a nucleus with $n$ neutrons and $z$ protons has been included in the
list of pseudo-observables accessible with the codes {\tt TOPAZ0} and 
{\tt ZFITTER}. In this way one can add the APV results in the LEP EWWG 
`global' electroweak fits, checking the corresponding effect when added 
to the existing precision measurements.
%\end{abstract}

\vspace{5cm}

\footnoterule

\vspace{1mm}

\noindent
$^{\dagger}$
{\small  
Work supported by the European Union under contract HPRN-CT-2000-00149.
}
\clearpage

%% file: APV_text.tex
\newcommand{\shq}{{\hat s}^4}
%---------------------
\section{Introduction}
%---------------------

Recently we have been asked to include atomic-parity violation (hereafter 
APV) parameters in the list of pseudo-observables (hereafter PO) that are 
accessible with the {\tt FORTRAN} codes 
{\tt TOPAZ0}~\cite{Montagna:1999kp} and {\tt ZFITTER}~\cite{Bardin:1999yd},
so to include the APV results in the LEP EWWG electroweak fits.

The reason for this operation is that there are now precise experiments 
measuring APV in cesium~\cite{cesium}, at the $0.4\%$ level, 
thallium~\cite{thallium}, lead~\cite{lead} and
bismuth~\cite{bismuth}. Moreover, according to~\cite{Blundell:1996qj}, the 
uncertainties associated with the atomic wave-functions have been reduced 
to another $0.4\%$ for cesium.
For additional uncertainties associated with the value of the tensor
polarizability we refer to~\cite{Groom:2000in}.
Note however that there is an intrinsic difference between the PO at the
$\zb$-resonance, e.g.\ $\gz$, $\sigma^0_{\had}$, $\afba{0}$ etc, and 
the APV parameters where the typical scale is dictated by the limit
of zero momentum transfer in the APV Hamiltonian. This fact alone
is the origin of a comparatively larger theoretical uncertainty which is
due to our basic ignorance of QCD corrections in this regime.

The investigation of APV has been the subject of a number of studies
made in the 80's by Marciano and Sirlin~\cite{Marciano:1983mm} 
and~\cite{Marciano:1984ss}.
For {\tt TOPAZ0}, which is based on the generalized minimal subtraction 
scheme~\cite{Passarino:1990ey}, it has been relatively simple to include
all recently computed higher-order effects in the old $\MSB$ calculation.
For {\tt ZFITTER} instead, the authors have been able to produce a novel
evaluation of the APV parameters in the on mass-shell (OMS) scheme.
The current value for the weak-charge is
%--
\bq
Q_{\sss{W}}({\rm Cs}) = -72.06 \pm 0.28 \pm 0.34\,({\rm theo.})
\eq
%--
For a recent evaluation of $Q_{\sss{W}}$ we refer, again, 
to~\cite{Groom:2000in} 
where the program {\tt GAPP}~\cite{Erler:1999ug} has been used.

%----------------------------------------------
\section{Upgrading the $\MSB$ calculation}
%----------------------------------------------
 
The electron--quark parity-violating Hamiltonian at zero momentum-transfer 
will be conventionally parametrized as follows:
%--
\bqa
H_{\sss {PV}} &=& 
\frac{\gf}{\sqrt 2}
\lpar
   C_{1u}{\bar e} \gadu{\mu}\gfd e {\bar u} \gadu{\mu} u
 + C_{2u}{\bar e} \gadu{\mu}     e {\bar u} \gadu{\mu} \gfd u
 + C_{1d}{\bar e} \gadu{\mu}\gfd e {\bar d} \gadu{\mu} d
 + C_{2d}{\bar e} \gadu{\mu}     e {\bar d} \gadu{\mu} \gfd d
\rpar,
\label{APV_MS}
\eqa
%---
where the ellipsis represents heavy-quark terms and we have factorized
 out the Fermi constant $\gf$.
In heavy atoms the dominant part of parity-violation is proportional to
the so-called weak-charge $Q_{\sss{W}}$
%--
\bq
Q_{\sss{W}}(Z,A)= 2\,\Bigl[\lpar Z+A\rpar\,C_{1u} + \lpar 2\,A-Z\rpar\,
C_{1d}\Bigr],
\eq
%--
We have taken the calculation by Marciano and Sirlin which is performed
in the modified minimal subtraction scheme ($\MSB$) and have extended it
to include all higher-order effects presently known. To summarize:
two-loop leading contribution for the $\rho$-parameter~\cite{Fleischer:1993ub},
exact $\ord{\alpha\als}$ corrections~\cite{Kniehl:1990yc}, 
$\ord{\alpha\als^2}$ corrections to $\rho$~\cite{Chetyrkin:1995js},
next-to-leading two-loop heavy top corrections~\cite{Degrassi:2000jd}.
At the same time, an attempt has been made to evaluate the theoretical 
uncertainty at the level of electroweak and of QCD corrections.

{\tt TOPAZ0} now returns, among all PO, the two quantities, $C_{1u}$ and
$C_{1d}$ of \eqn{APV_MS}. They are defined as follows:
%---
\bqa
C_{1u} &=& -\frac{1}{2}\,\rho'_{\rm PV}\,\biggl[1 - \frac{8}{3}\,
\kappa'_{\rm PV}(0)\sin^2{\hat\theta}(\mws)\biggr],  \nl
C_{1d} &=& \frac{1}{2}\,\rho'_{\rm PV}\,\biggl[1 - \frac{4}{3}\,
\kappa'_{\rm PV}(0)\sin^2{\hat\theta}(\mws)\biggr],
\label{factorized}
\eqa
%---
where $\sin^2{\hat\theta}(\mws)$ is the $\MSB$ weak-mixing angle at the scale
$\tHs = \mw$. We adopt a specific implementation of the re-summation
procedure where the pair $\mw$ and $\sin^2{\hat\theta}(\mws)$ is the
solution of a system of coupled non-linear equations that include all
available higher-order effects, as described in Sect. 6.11 and 8 of 
ref.~\cite{Bardin:1999ak}. Moreover,
%---
\bqa
\rho'_{\rm PV} &=& \rho - \frac{\alpha}{2\,\pi}\,\Biggl[
1 + \frac{1}{\shs} + 4\,\hatv{e}\,B_{\rm p(np)} + \frac{9}{16\,\shs\chs}
\lpar 1 - \frac{16}{9}\shs\rpar\lpar 1 + \hatv{e}^2\rpar\Biggr],  \nl
\kappa'_{\rm PV}(0) &=& \kappa_{\rm PV}(0) - \frac{\alpha}{2\,\pi\shs}\,
\Biggl[ \frac{9-8\shs}{8\,\shs} - \frac{\hatv{e}}{6}\,\lpar
\ln\frac{\mzs}{\mes} + \frac{1}{6}\rpar + \lpar \frac{9}{4} - 4\shs\rpar\,
\hatv{e}\,B_{\rm p(np)} \nl
{}&&+ \frac{9}{16\,\shs\chs}\,\lpar \frac{1}{2}
\hatv{e} + \frac{16}{9}\shq\rpar\,\lpar 1 + \hatv{e}^2\rpar\Biggr],
\label{APV_parameters}
\eqa
%--
where $\hatv{e} = 1 - 4\,\shs$, $\shs = \sin^2{\hat\theta}(\mws)$ and where
we have two different treatments of the $\zb$--$\ph$ boxes --- 
perturbative~\cite{Marciano:1983mm}
%--
\bq
B_{\rm p} = \ln\frac{\mzs}{m^2} + \frac{3}{2}, \quad m = \muq = \md = 75\,\MeV,
\label{APV_boxes}
\eq
%--
and non-perturbative~\cite{Marciano:1984ss}
%--
\bqa
B_{\rm np} &=& K + \frac{4}{5}\,\lpar\xi_1\rpar^p_B,  \nl
K &=& \mzs\,\int_{M^2}^{\infty}\, \frac{du}{u(u+\mzs)}\,\Biggl[ 1 -
\frac{\als(u)}{\pi}\Biggr], \quad  \lpar\xi_1\rpar^p_B = 2.55,
\label{splitbox}
\eqa
%--
where $M$ is a mass scale representing the onset of the asymptotic behavior,
i.e. the regime where $\als$ becomes small. We observe a plateau of
stability in $K$ for $M$ centered around $0.5\,$GeV and this is the numerical
value used. Furthermore we used the following form for the $\rho,\kappa$
parameters~\cite{Marciano:1983mm}:
%--
\bqa
\rho &=& 1 + \frac{\alpha}{4\,\pi\shs}\,\Biggl[\frac{3}{4\,\shs}\,\ln\chs -
\frac{7}{4} + \frac{3}{4}\,\frac{\mts}{\mws}\,
\lpar 1 + \delta_{\rm EW} + \delta_{\rm QCD}\rpar 
\nl
{}&&+ \frac{3}{4}\,h\,\lpar
\frac{\ln(\chs/h)}{\chs-h} +\frac{1}{\chs}\,\frac{\ln h}{1-h}\rpar\Biggr],
\nl
\kappa_{\rm PV}(0) &=& 1 - \frac{\alpha}{2\,\pi\shs}\,\Biggl[ \frac{7}{9} -
\frac{\shs}{3} +\frac{\qf}{3}\,\asums{f}\lpar \tcie -2\,\qf\shs\rpar\,
\ln\frac{\mfs}{\mws}\Biggr].
\label{again_parameters}
\eqa
%--
In the previous equation we have $h = \mhs/\mzs$.
The strange quark mass is effectively chosen to be $\ms = 250\,$MeV so that
(with effective $\muq = \md = 75\,\MeV$) we recover the dispersive analysis for
the $\zb$--$\ph$ transition where $\Pi_{{\sss{Z}}\ph}(0)$ is rewritten in 
terms of a dispersion relation with the kernel connected to $\sigma(e^+e^-\to 
{\rm hadrons})$.
Inside \eqn{again_parameters} $\delta_{\rm EW(QCD)}$ are the LO+NLO
electroweak $\Bigl(\ord{\als^2+\als^3}\mbox{~QCD}\Bigr)$ correction to 
$\rho$. The evaluation of $\rho$ and $\shs$ includes the best available LO+NLO 
terms~\cite{Bardin:1999ak}.

{\tt TOPAZ0} default is the perturbative formulation of the {\em factorized} 
result of \eqn{factorized}. There is the option of using some 
{\em additive} formulation where
%---
\bqa
C_{1u} &=& -\frac{1}{2}\,\rho\,\biggl[1 - \frac{8}{3}\,
\kappa_{\rm PV}(0)\sin^2{\hat\theta}(\mws)\biggr] + \Delta_u\,,  
\nl
C_{1d} &=& \frac{1}{2}\,\rho\,\biggl[1 - \frac{4}{3}\,
\kappa_{\rm PV}(0)\sin^2{\hat\theta}(\mws)\biggr] + \Delta_d\,,
\label{additive}
\eqa
%---
where $\Delta_{u,d}$ are obtained from \eqn{factorized} by expanding and
by neglecting terms of $\ord{\alpha^2}$.
%--
%----------------------------------------------
\section{Atomic parity-violation in OMS scheme}
%----------------------------------------------

The old result of~\cite{Marciano:1983mm} has been completely re-derived in the
OMS scheme. Here, the technical problem is represented by the extraction of 
the limit of zero momentum transfer from the expressions that have been
derived for the process $ee\to t\bar{t}$ \cite{Bardin:2000kn}. Here the process
under consideration is the $\tman$-channel scattering $ee\to uu$ and what
we need is naturally contained in the results of Ref.~\cite{Bardin:2000kn} 
since they were derived retaining all masses and, therefore, the limit of zero
momentum transfer, $Q^2 << $ (all) $m^2$ is possible. 

It is rather easy to take the limit $Q^2\to 0$
for vertices and self-energy functions since they depend only on this 
variable.
For boxes the procedure is more complex due to their complicated dependence on 
$\sman$ and $\tman$ invariants. Fortunately enough, $\zb\zb$ and $\zb\ph$ 
boxes form a gauge invariant sub-set of the whole result and for $\wb\wb$ 
boxes one has to replace, in the corresponding limit, only the $\gpar=1$ part
of the result, which is well defined and simple. This fact triggered the 
strategy for a calculation where we take all contributions but boxes from 
the $Q^2\to 0$ limit of the $ee\to t\bar{t}$ form factors and were we have 
re-computed, from scratch, box diagrams at $Q^2=0$.
Note that this calculation was done with the aid of the computer system
described in Ref.~\cite{CalcPHEP:2000}. 

For our calculation we compare the APV Hamiltonian of \eqn{APV_MS}
with its $ee\to t\bar{t}$ analog, Eq.(I.10) of~\cite{Bardin:2000kn}: 
%---
\bqa
{\cal A}_{\sss{\zb}} \lpar 0 \rpar &=& 
%{\cal A}^{\sss{OLA}}_{\sss{\zb}}&=&
%\ib\gspi\,e^2\,4
\tcie\tcif\frac{\pi\alpha}{\siws \cows (-\mzs)}
\Biggl\{\gadu{\mu}{\gdp } \otimes
        \gadu{\mu}{\gdp } \vverti{}{\sss{LL}}{0} +d_e\gadu{\mu}     
      \otimes \gadu{\mu} {\gdp}\vverti{}{\sss{QL}}{0} 
\nll &&
+d_f\gadu{\mu}{\gdp}\otimes\gadu{\mu}\vverti{}{\sss{LQ}}{0}
+d_e d_f \gadu{\mu} \otimes\gadu{\mu}\vverti{}{\sss{QQ}}{0}
\Biggr\}.
\label{APV_eett}
\eqa
%---
Here $(0)$ stands for $Q^2=0$ and we write only one argument since 
box contributions are excluded. Moreover, 
%---
\bqa
\gdp=1+\gfd\,,\qquad d_f &=& - 4 |Q_f| \siws\,.
\eqa
%---
From \eqnsc{APV_MS}{APV_eett}
we immediately derive a relation between the APV parameters $C_{1f}$ and 
$C_{2f}$ and the $ee\to t\bar{t}$ form factors at zero momentum transfer:
%---
\bqa
C_{1f} &=&  
I^{(3)}_f \Bigl[f_{\sss{LL}} + d_f f_{\sss{LQ}}-\Delta r\lpar 1+d_f\rpar\Bigr],
\nll 
C_{2f} &=& 
I^{(3)}_f \Bigl[f_{\sss{LL}} + d_e f_{\sss{QL}}-\Delta r\lpar 1+d_e\rpar\Bigr].
\eqa
%---
Here $f=u,d$ and
%---
\bq
f_{\sss{LL,QL,LQ}}= 1 + \frac{\alpha}{4\pi\siws} F_{\sss {LL,QL,LQ}}(0).
\eq
%---
After a lengthy but straightforward calculations, we are able to reproduce
the following generic expressions:  
%---
\bqa
C_{1u} &=& -2 \tcie \rho_{\sss{PV}}\lpar I^{(3)}_{u}
                -2\qu\kappa_{\sss {PV}} \siws \rpar
\nll &&
            + \frac{\alpha}{\pi} \Biggl[
              Q_e^2 a_e v_u
            +\frac{1}{3} Q_u Q_{\nu} \lpar \ln r_{{\sss W}e} + \frac{1}{6} \rpar
            +\frac{2}{3} Q_u  Q_e v_e a_e 
                      \lpar \ln r_{{\sss Z}e}+ \frac{1}{6} \rpar
\nll &&
            +C^{\sss{WW}}_f
            +3 Q_u a_u Q_e v_e  \lpar \ln r_{{\sss Z}u} +\frac{3}{2} \rpar
            +\frac{3}{4\siws\cows} v_u a_u \lpar v_e^2+a_e^2 \rpar \Biggr],
\label{APV_generic}
\eqa
%---
where we introduced a notation $r_{ij}=\minds{i}/\minds{j}$ and a fictitious 
term with non-zero neutrino charge in order
to have a completely general representation and where
\clearpage
%---
\bqa
C^{\sss{WW}}_f=
     \lcbr
\begin{array}{l}
 \;\; \ds{\frac{1}{2\siws}} \quad\, \mbox{~for~} f=u\,, \\[5mm]
      \ds{-\frac{1}{8\siws}}\quad   \mbox{~for~} f=d\,,
\end{array}
     \right.
\eqa
%---
is a contribution, originating from the $\wb\wb$ box, which is different 
for $u$ and $d$ channels (direct--crossed). 

The other APV parameters can be obtained with the aid of some simple 
substitutions:
%--
\bq
C_{2u}    = C_{1u}   \bigg|_{\ds e \leftrightarrow u, Q_\nu \rightarrow Q_d}\,,
\qquad
C_{1(2)d} = C_{1(2)u}\bigg|_{\ds u \leftrightarrow d}\;.
\eq
%--
The terms of the second and third rows of \eqn{APV_generic} are identical 
to corresponding terms of the $\MSB$ result. In the sequential order they 
are due to: QED vertex in $\zb$ exchange, $\wb$ abelian vertex in $\ph$ 
exchange (with neutrino charge), $\zb$ abelian vertex in $\ph$ exchange; 
$\wb\wb$, $\zb\ph$ and $\zb\zb$ boxes.

The only difference with respect to the $\MSB$ result is present in the first 
term. The factor $\rho$ is almost the same: 
%--  
\bq 
\rho_{\sss {PV}} = 
        1 + \frac{\alpha}{4\pi\siws}  \Biggl\{ \frac{3}{4}
  \Biggl[ - \frac{1}{\siws}\ln\cows - \frac{\rhw}{1-\rhw}\ln\rhw 
          + \frac{\rhw}{1-\rhz} \ln\rhz \Biggr]-\frac{7}{4}
          - \Delta \rho^{\fer}(0) \Biggr\}, 
\label{phoAPV}
\eq
%--
where we use instead the full expression for $\rho^{\fer}(0)$,
%--
\bq
   \Delta \rho^{\fer}(0) = \frac{\Sww^{\fer}(0)-\Szz^{\fer}(0)}{\mws}\;,
\eq
%--
contrary to the approximation made above where only the (leading) quadratic 
term in $\mt$ is retained. 
The difference, being proportional to light fermion masses is numerically 
rather small.

However, the main difference with the $\MSB$ calculations is confined in 
the APV parameter $\kappa_{\sss {PV}}$ for which, in the OMS-scheme,
we derived:
%--
\bq
 \kappa_{\sss {PV}} = 
% ! -2d0/3d0*ve*(Lnmuzm-Lnmuem+1d0/6d0)
 1 + \frac{\alpha}{4\pi\siws}
 \Biggl\{ \lpar \frac{1}{6}+7\cows \rpar L_\mu(\mws)
      -\frac{8}{9}-\frac{2}{3}\cows
%\nll &&
             -\frac{\cows}{\siws} 
       \lpar \Delta \rho^{{\bos},F}  + \Delta \rho^{{\fer},F} \rpar
      -\Pi^{\fer}_{{\sss Z}\gamma}(0) \Biggr\},
\eq
%--
where $L_\mu(\mws) = \ln{\ds\frac{\mws}{\tHss}}$.
The gauge invariant Veltman $\Delta \rho$ parameter is
%--
\bq
   \Delta \rho = \frac{\Sww^{\fer}(\mws)-\Szz^{\fer}(\mzs)}{\mws}\;,
\eq
%--
and contains both the {\em bosonic} and the {\em fermionic} components. 
We explicitly give the bosonic part, $\Delta \rho^{\bos}$ (for definition
of finite part $\sfbff{0}$ of $\sbff{0}$ functions see \cite{Bardin:1999ak}):
%---
\bqa
\Delta \rho^{{\bos},F}
     &=&
   \lpar \frac{1}{ 12 \cowf}+\frac{4}{3\cows}
              -\frac{17}{3}-4\cows    \rpar
\biggl[  \fbff{0}{-\mws}{\mw}{\mz}
                  -\cows       \fbff{0}{-\mzs}{\mw}{\mw} \biggr]
\nll &&
+\lpar 1-\frac{1}{3}\rhw
 +\frac{1}{12} \rhw^2  
 \rpar                         \fbff{0}{-\mws}{\mw}{\mh}
\nll &&
-\lpar 1 - \frac{1}{3} \rhz
         + \frac{1}{12}\rhz^2 \rpar \frac{1}{\cows}
                                   \fbff{0}{-\mzs}{\mz}{\mh} 
 -4 \siws                          \fbff{0}{-\mws}{\mw}{0}
\nll && 
 + \frac{1}{12} \Biggl[ 
          \lpar 
 \frac{1}{\cowf} + \frac{6}{\cows} - 24 + \rhw \rpar 
               {L}_{\mu}(\mzs)
       + \siws \rhw^2 \lrbr  {L}_{\mu}(\mhs) - 1 \rrbr
\nll &&
    -\lpar \frac{1}{\cows} + 14 + 16 \cows - 48 \cowf + \rhw \rpar 
               {L}_{\mu}(\mws)
    -\frac{1}{\cowf} - \frac{19}{3 \cows} + \frac{22}{3}
         \Biggr].
\eqa
%---
To establish a link with the $\MSB$ calculation we introduce the usual notion 
of leading and reminder terms:
%---
\bqa
 \kappa_{\sss {PV}}  & = &
% ! -2d0/3d0*ve*(Lnmuzm-Lnmuem+1d0/6d0)
 1 + \frac{\alpha}{4\pi\siws}
 \biggl\{  \lpar \Delta \kappa_{\sss {PV}}\rpar_{\rm lead} 
         + \lpar \Delta \kappa_{\sss {PV}}\rpar_{\rm rem }\biggr\}\,,
\eqa
%---
where the leading term contains only $\Delta \rho$ and the reminder
contains all the rest:
%---
\bqa
\lpar \Delta \kappa_{\sss {PV}}\rpar_{\rm lead}& = &
             -\frac{\cows}{\siws} 
\lpar \Delta \rho^{{\bos},F}  + \Delta \rho^{{\fer},F} \rpar\bigg|_{\tHs=\mw}\,,
\nll 
\lpar \Delta \kappa_{\sss {PV}}\rpar_{\rm rem} & = &
%  \lpar \frac{1}{6}+7\cows \rpar L_\mu(\mws)
      -\frac{8}{9}-\frac{2}{3}\cows
      -\Pi^{\fer}_{{\sss Z}\gamma}(0)\bigg|_{\tHs=\mw}\,.
\eqa
%---
Numerically, $\lpar\Delta\kappa_{\sss {PV}}\rpar_{\rm lead}$ and 
$\lpar \Delta \kappa_{\sss {PV}}\rpar_{\rm rem }$ are nearly equal and one 
might think that the usual leading--reminder splitting,
the standard factorization of contributions with different scales 
and re-summation (see~\cite{Bardin:1999ak}),
%--
\bq
 \kappa_{\sss {PV}} = 
 \lrbr 1 + \frac{\alpha}{4\pi\siws}
               \lpar \Delta \kappa_{\sss {PV}}\rpar_{\rm lead} \rrbr
 \lrbr 1 + f_c \frac{\alpha}{4\pi\siws}
               \lpar \Delta \kappa_{\sss {PV}}\rpar_{\rm rem}  \rrbr,
\label{kappa_factorized}
\eq
%--
with a conversion factor
\bqa
f_c=\frac{\sqrt{2} G_{\mu} \mzs \siws \cows}{\pi\alpha}\,,
\eqa
%---
is not too well justified for the APV parameter $\kappa_{\sss {PV}}$.
Note, however, that the factorized form of \eqn{kappa_factorized} is fully 
consistent with the $\MSB$ result \eqn{again_parameters} if we identify
%--
\bq
\sin^2{\hat{\theta}}_{\sss{W}}(\mw)=\lrbr 1 - f_c \frac{\alpha}{4\pi}
               \frac{\cows}{\siwf}
               ~\Delta \rho^{F}\bigg|_{\tHs=\mw} \rrbr\siws\,.
\eq
%--
As done before, for $\Pi^{\fer}_{{\sss Z}\gamma}(0)$ we use effective 
quark masses which are consistent with a dispersive treatment of 
$\Pi^{\fer}_{{\sss Z}\gamma}$ at zero scale.

%two-loop leading contribution for the $\rho$-parameter~\cite{Fleischer:1993ub},
%exact $\ord{\alpha\als}$ corrections~\cite{Kniehl:1990yc}, 
%$\ord{\alpha\als^2}$ corrections to $\rho$~\cite{Chetyrkin:1995js},

Finally, we apply mixed QCD $\Bigl(\ord{\als^2+\als^3}\Bigr)$
and LO+NLO electroweak two-loop corrections for
Veltman $\Delta \rho$ parameter. For $\rho_{\sss {PV}}$ we stick with one-loop 
(non-re-summed) result \eqn{phoAPV}, since there the notion 
of leading--reminder 
splitting fails completely (numerically it looks like $+3.5 - 3.0 = 0.5$). 
For the same reason, we apply to $\Delta\rho^{\fer}(0)$ only the mixed QCD
but not the electroweak two-loop corrections, as already done in
the first \eqn{again_parameters}. 
The latter, as well as the other electroweak NLO corrections for remainder 
terms, although not implemented are successively used to evaluate the 
theoretical uncertainty in the electroweak sector of the OMS scheme.

%---------------------------------------
\section{Theoretical uncertainty in APV}
%---------------------------------------

In order to discuss the present level of theoretical uncertainty in
atomic parity-violation we start with the $\MSB$ results
for $Q_{\ssW}({\rm Cs})$ that are shown in \tabn{tabmsb},
corresponding to $\mz= 91.1875\,$GeV, $\mh = 150\,$GeV and 
$\als(\mzs) = 0.119$. {\tt ZFITTER} numbers corresponding to 
`Add/Pert' setup are added to the third row of the Table.
%--
\begin{table}[hpt]\centering
\begin{tabular}{|c|c|c|c|}
\hline
\hline
$\mt\,$[GeV]  &   170      & 175      & 180       \\
\hline
%             &            &          & \\
Fact/Pert     &   -72.9712 & -72.9632 & -72.9551  \\
%             &            &          & \\
\hline
Fact/Non Pert &   -73.1994 & -73.1932 & -73.1869  \\
%             &            &          & \\
\hline
Add/Pert      &   -72.9732 & -72.9658 & -72.9582  \\
{\tt ZFITTER} &   -72.9762 & -72.9698 & -72.9637  \\ 
\hline
Add/Non Pert  &   -73.2026 & -73.1969 & -73.1912  \\
%             &            &          & \\
\hline
\hline
\end{tabular}
\caption[]{Predictions for $Q_{\sss{W}}({\rm Cs})$ from {\tt TOPAZ0} and
{\tt ZFITTER} for $\mz= 91.1875\,$GeV, $\mh = 150\,$GeV and 
$\als(\mzs) = 0.119$.}
\label{tabmsb}
\end{table}
%--
The $\mt$-dependence of $Q_{\sss{W}}({\rm Cs})$ is shown in \tabn{tabmsb}
where we register a $0.22(0.17)$ per-mill increase for $\mt$ between
$170\,$GeV and $180\,$GeV and for Pert(Non Pert). As for the $\mh$ 
dependence we have computed a decrease of about $0.7$ per-mill for
$\mh$ between $150\,$GeV and $300\,$GeV.

The associated theoretical uncertainty is approximately $3.2$ per-mill and
it is largely dominated by QCD effects. Let us consider the main sources of 
uncertainty. In the original calculation of Marciano and Sirlin we have
a dependence of the result on light quark masses. 
This appearance can be seen in \eqn{APV_parameters} and in the perturbative
treatment of boxes, \eqn{APV_boxes}.

From 1983 the accuracy
associated to the weak-charge $Q_{\sss{W}}$ has been considerably reduced
and we cannot include it in the list of high-precision PO 
if the result contains logarithmic enhancements due to light quark masses.

In their second paper Marciano and Sirlin~\cite{Marciano:1984ss}
have suggested how to go beyond the partonic-language.
One should distinguish quark masses in the $\zb$--$\ph$ transition
and in $\zb$--$\ph$ boxes. Light quark masses, $m_{uds}$, are then fixed
to parameterize the dispersive result for the $\zb$--$\ph$ transition and
are not varied anymore in evaluating the theoretical uncertainty.

Furthermore we have $\zb$--$\ph$ box diagrams where quark masses show up as 
the consequence of the zero momentum transfer limit. Here, according to
the suggestion of~\cite{Marciano:1984ss} we split the boxes into a
low-frequency part, approximated with the Born contribution for a 
physical nucleon (the $\lpar\xi_1\rpar^p_B$ term in \eqn{splitbox}), and
an high-frequency part (the $K$-term in \eqn{splitbox}) that includes 
$\ord{\als}$ corrections where light quark masses disappear. The mass scale 
$M$ separating low- from high-frequency parts is, of course, arbitrary and 
only subjected to the requirement that $\als(Q^2)$ starts to become small
for $\mid Q^2\mid > M^2$ and that $M > \Lambda_{\rm QCD}$.
However, with the most complete evaluation of $\als$ (up to three loops)
we have found a plateau of stability for the result, i.e.
for $M$ between $0.5$ and $0.6(0.8)$ $K$ goes from $9.2016$ to
$9.1737(8.7818)$ and $Q_{\sss{W}}$ has a variation of $0.02(0.3)$ per-mill.
Therefore we fix $0.5 \le M \le 0.6$.

Instead of varying light quark masses between undefined limits we prefer to
estimate the theoretical uncertainty by comparing the perturbative result
with light quark masses fixed to reproduce the dispersive approach to the
$\zb$--$\ph$ transition with a non-perturbative anzatz based on a low-frequency
high frequency splitting at a mass scale of about $0.5\,$GeV.
Note that when comparing $B_{\rm np}(M)$ with the perturbative factor
$\ln\mzs/M^2 + \frac{3}{2}$ we find that the perturbative
approach overestimates the effect of about $5.6\%(1.9\%)$ at $M = 
0.5(0.8)\,$GeV.

Furthermore, the differences in the factorized~\eqn{factorized} versus 
additive~\eqn{additive} formulation of the coefficients $C_{1u,1d}$ is
approximately $0.05$ per-mill signaling that, from {\tt TOPAZ0}'s treatment
alone, pure electroweak higher orders are relatively under control. Another 
way of testing the electroweak theoretical uncertainty is, as usual, to 
compare two different renormalization schemes with the same input parameter 
set. When we compare {\tt ZFITTER} in the preferred setup with
{\tt TOPAZ0} additive/perturbative we obtain a relative difference of
$0.04(0.05,0.08)$ per-mill at $\mt= 170(175,180)\,$GeV. 
However, an internal evaluation of electroweak theoretical uncertainties 
within {\tt ZFITTER} (realized by evaluating the effect of the
electroweak two-loop corrections which are not included in the preferred setup)
shows a value of about $\pm 0.25$ per-mill.
Again, the conclusion is that theoretical uncertainty is completely dominated 
by QCD effects at zero momentum transfer.

Finally, let us define an effective APV weak-mixing angle by the following
relation:
%--
\bq
\sin^2\theta_{\rm APV} = \kappa'_{\rm PV}(0)\sin^2{\hat\theta}(\mws).
\eq
%--
For $\mz= 91.1875\,$GeV, $\mh = 150\,$GeV and $\als(\mzs) = 
0.119$ we obtain 
$\sin^2\theta_{\rm APV} = 0.231601$ $(0.232123)$,
corresponding to perturbative (non-perturbative) treatment.
\vspace{0.5cm}

%% file: expansions.tex
%--------------------------------
{\bf Appendix: Taylor expansions}
%--------------------------------
Here we list all expansions that are needed in order to reproduce the OMS
results. Note that we need at most terms of $\ord{\sman}$:
%---
\bqa
&& \cff{0}{-\ums}{-\ums}{-s}{\um}{M}{\um}
  = \frac{1}{M^2}\Biggl\{
     -1 - \frac{7}{2}\ruM - \frac{37}{3}\ruM^2 
   - \lpar 1 + 3\ruM + 10 \ruM^2 \rpar \ln\ruM   
\nll &&  \hspace{6mm}
     +\frac{s}{M^2} 
  \Biggl[
\frac{1}{6 \ruM}+\frac{13}{12}+\frac{52}{9}\ruM+\frac{673}{24}\ruM^2   
      +\lpar\frac{1}{2}+\frac{10}{3}\ruM+\frac{35}{2}\ruM^2 \rpar \ln\ruM  
\Biggr] 
 \Biggr\},
\eqa
%---
where $M = \mz,\mh$ and we remind a short hand notation for mass ratios:
$r_{ij}=\ds\frac{m^2_i}{m^2_j}\,$.

The other expansions read:
%---
\bqa
&&\hspace{-5mm} 
\cff{0}{-\um^2}{-\um^2}{-s}{\dm}{\mw}{\dm} = \frac{1}{\mws}
\Biggl\{ -1-\rdw-\rdw^2 - \lpar\frac{5}{2}+8\rdw\rpar\ruw-\frac{10}{3}\ruw^2
\nll && \hspace{65mm}
-\Biggl[1+2\rdw+3\rdw^2+\lpar 1 + 6\rdw \rpar\ruw + \ruw^2\Biggr] \ln\rdw 
\nll && \hspace{1mm}
 +\frac{s}{\mws}\Biggl(
 \frac{1}{6}\Biggl[\frac{1}{\rdw}+\frac{11}{2}+ 13\rdw +\frac{47}{2}\rdw^2
 + \lpar \frac{1}{\rdw} + \frac{62}{3}+97\rdw \rpar\ruw
 + \lpar \frac{1}{\rdw} + \frac{187}{4}\rpar\ruw^2 
\nll && \hspace{15mm}
 + \frac{1}{\rdw}\ruw^3
\Biggr]
 +\Biggl[
\frac{1}{2} + 2\rdw + 5\rdw^2 + \lpar \frac{4}{3} + 10 \rdw \rpar\ruw 
  + \frac{5}{2} \ruw^2 \Biggr] \ln\rdw  
\Biggr)
\Biggr\},
\eqa
%---
\bqa
\cff{0}{-\ums}{-\ums}{-s}{\mw}{0}{\mw}=
  \frac{1}{\mws } \Biggl[1+\frac{\ruw}{2}+\frac{\ruw^2}{3}
 +\frac{s}{6\mws} \lpar \frac{1}{2}+\frac{1}{3} \ruw \rpar \Biggr],
\eqa
%---
\bqa
 \cff{0}{0}{0}{0}{\mh}{0}{\mz}=-\frac{1}{\mzs-\mhs} \ln\rhz\,,
\eqa
%---
\bqa
\fbff{0}{-\um^2}{\mw}{0} = - {L_{\mu}(\mws)}
                 + 1 + \frac{1}{2} \ruw+\frac{1}{6} \ruw^2\,,
\eqa
%---
\bqa
\fbff{0}{-\um^2}{M}{\um} = - {L_{\mu}(M^2)}
          + \lpar \ruM + 2\ruM^2 \rpar\ln\ruM
      + 1 + \frac{1}{2}\ruM+\frac{5}{3}\ruM^2\,,
\eqa
%---
\bqa
\bff{0p}{-\um^2}{M}{\um} &=& -\frac{1}{2M^2}\,,
\nll[1mm]
\fbff{0}{-s}{m}{m} &=& - {L_{\mu}(m^2)} + \frac{s}{6 m^2}\,,
\eqa
%---             
\bqa
\fbff{0}{-s}{m}{M}&=&
   1 + \lrbr M^2 {L_{\mu}(M^2)} - m^2 {L_{\mu}(m^2)} \rrbr \frac{1}{m^2-M^2}
\nll[1mm] && \hspace{3mm}
   + s \Biggl[ \frac{m^2+M^2}{2(m^2-M^2)^2} 
         -\frac{m^2 M^2}{(m^2-M^2)^3}\ln\lpar\frac{m^2}{M^2}\rpar\Biggr].
%      + s^2 (\lrbr \lpar m^2+mp^2 \rpar^2+8 m^2 mp^2 \rrbr S_x(m,mp)^4/6
%                -m^2 mp^2 (m^2+mp^2) ln(m^2/mp^2) S_x(m,mp)^5);
\eqa
%---